# Strongly connected components

-Algorithm for finding the strongly connected components of a graph-

Author: Vlad-Andrei Munteanu

## 1. Abstraction

A directed graph G (V, E) is strongly connected if and only if, for any pair of vertices X and Y from V, there exists a path from X to Y and a path from Y to X. In Computer Science, the partition of a graph in strongly connected components is represented by the partition of all vertices from the graph, so that for any two vertices, X and Y, from the same partition, there exists a path from X to Y and a path from Y to X and for any two vertices, U and V, from different partition, the property is not met. The algorithm presented below is meant to find the partition of a given graph in strongly connected components in O (numberOfNodes + numberOfEdges * log* (numberOfNodes)), where log* function stands for iterated logarithm.

## 2. Algorithm

For a better understanding of the algorithm it is recommended a revision of Disjoint-set Data Structure (section 6) and DFS oriented graph traversing (section 7). We will consider a strongly connected component as a set. With no doubt, two strongly connected components represent two disjoint sets because, if there exists a vertex X which belongs to the first strongly connected component as well as to the second strongly connected component, it means that for any vertex U from the first strongly connected component and any vertex V from the second strongly connected component, we can find both, a path from V and a path from U to V through X. This holds because X belongs to the same strongly connected component as U and V. From here, it results that the two strongly connected components we have selected can be joined in a bigger one. Using induction, we can prove that, if two or more strongly connected components are linked by one or more vertices, that means that they can be jointed. Therefore, the strongly connected components of a graph are pairwise disjointed. Initially, we will consider that every vertex is in its own strongly connected component (by definition, a vertex is a strongly connected component) and, using a DFS traversal, we will link those vertices that are in the same strongly connected component). The Algorithm is recursive type, and in the following lines, I will present the steps that must be followed:

1. the X vertex is added in the DFS stack and it is marked as being visited (either as DFS source, either as adjacent vertex to the top vertex of the stack);

2. iterate through all adjacent vertices of X in any order and for every neighbor, Y, it will be:

- the case in which Y had already been visited, so the algorithm just verifies if any vertex from the stack is in the same strongly connected component as Y, and, if case positive, it joins the strongly connected components of X and Y;

- the case in which the neighbor has not been visited yet, so it must be introduced in the stack. The algorithm is executed for the vertex Y, starting from the first step. After the execution of the algorithm for Y, it is popped out from the stack and, now being visited, the checking mentioned above between the nodes in the stack and Y needs to be done. If case positive, the strongly connected components of X and Y must be joined as above.

3. the vertex X is popped out from the stack;

A crucial point of the algorithm is how we may know that a vertex from the stack is in the same connected components as the neighbor. In order to solve this problem, we must hold the minimum level from the DFS stack of each strongly connected component, which will be initialized with infinity. If the level of the strongly connected component of Y is infinity, there is no vertex in the stack which belongs to it. We have to reset to infinity the level of a strongly connected component when the vertex with the lowest level which belongs to that component is to be popped out from the DFS stack. When two strongly connected components are joined, the dominant one will hold the minimum level between its level and the level of the other one.

## 3. Proof of correctness

The last paragraphs focused on the algorithm itself and on the most important observation which allows us to use Disjoint-set Data Structure. The following paragraphs will focus on proving the correctness of the algorithm. We will take into account all the situations which could appear with the adding of a new vertex in the stack (let`s assume that the new added vertex is X):

-Case 1: The vertex X is the first pushed to the stack and the algorithm begins.

-Case 2: The vertex X has a father, vertex Y, and they belong to the same strongly connected component.

-Case 3: The vertex X has a father, vertex Y, and they belong to different strongly connected components.

Case 1: If X is the first pushed vertex, it does not have a father, so this will not be an attempt to change the strongly connected component (joining attempt is done only for the pairs {father, vertex}.

Case 2: If the X vertex has a father, node Y and they are found in the same strongly connected component, it means that there exists a path from X to Y (the path from Y to X being a direct edge). We will consider the path as X->$Vertex_1$->$Vertex_2$->$Vertex_3$->…->$Vertex_k$->Y (in the case in which there are more valid paths, consider the first which will be taken into account by the DFS traversal). Due to the properties of the DFS traversal which guarantees that all the paths from the current node which have not been visited until then will be taken into consideration, we could jump to the conclusion that if the vertex $Vertex_i$ is not visited, then all vertices $Vertex_i$, $Vertex_{(i+1)}$,…, $Vertex_k$ and Y are visited and they are in the DFS stack (if at least one from these is not visited it means that the chosen path is not the first one selected by the DFS traversal, and if they would not be in the stack and but visited, it would mean that a complete DFS traversal has been executed, so X, being previously visited too, cannot be added now) and, moreover, any vertex Z, which belongs to a path from X to Y, but does not belong to the chosen path, it is not visited (in the case in which it was previously visited, either a complete execution of DFS traversal has been executed, which leads to the contradiction discussed before, or the chosen path is not the first one -the first path being the one which goes through Z-, which leads to a contradiction of the hypothesis). Therefore, there exists a vertex, $Vertex_j$, with the property that all vertices $Vertex_j$, $Vertex_{(j+1)}$, …, $Vertex_k$, Y are visited and in the stack and all vertices X, $Vertex_1$, $Vertex_2$,…, $Vertex_{(j-1)}$ are not visited. From the same property of the DFS traversal, it is guaranteed that it must lead to the traversal of X, $Vertex_1$, $Vertex_2$, …, $Vertex_{(j-1)}$ at some point. Arrived in $Vertex_{(j-1)}$, we will find $Vertex_j$ which is visited and in the stack, so the strongly connected components of $Vertex_{(j-1)}$ and $Vertex_j$ are joined. Back from recursion to $Vertex_{(j-2)}$, it is confirmed the fact that $Vertex_{(j-1)}$ it is in the stack through the link with $Vertex_j$. Inductively, the link would be made between $Vertex_{(j-2)}$ and $Vertex_{(j-3)}$, between $Vertex_{(j-3)}$ and $Vertex_{(j-4)}$, …, between $Vertex_1$ and X, and, finally, between X and Y, being in the stack through $Vertex_j$. In conclusion, if two vertices X and Y which are connected by an edge are in the same component, the algorithm will join them.

Case 3: If the vertex X has a father, vertex Y, and they are in different strongly connected components, it means that there exists no path from X to Y (the path from Y to X exists through direct edge). As there exists no path from X to Y, it means that for any vertex Y` with the property that there exists a path from Y` to Y and vertex X` with the property that there exists a path from X to X`, there exists no path from X` to Y`. Therefore, the DFS traversal, would not find any path from any "son" of Y to any "father" of X, and so, no vertex which is in the same set as Y ( Y included) will be connected with no vertex which is in the same set as X( X included).

In conclusion, the behavior of the algorithm is as desired in all possible cases, so the correctness is demonstrated.

## 4. Complexity

The complexity is given by the DFS traversal and by a possible connection between of any two vertices which are connected through and edge, so we have O(numberOfNodes+numberOfEdges + numberOfEdges * log* (numberOfNodes)), which is O(numberOfNodes + numberOfEdges * log* (numberOfNodes)), where log* stands for the iterated logarithm.

## 5. Implementation

A simple and clean implementation of the algorithm can be done through five methods (you can see the C++ code below):

1. The method which generate the set of a specific element (vertex)

```cpp
int getComponent (int currentElement) {
   int currentKing = currentElement;
   while (currentSCC[currentKing] != currentKing) {
      currentKing = currentSCC[currentKing];
   }
   while (currentElement != currentKing) {
      int copyOfElement = currentElement;
      currentElement = currentSCC[copyOfElement];
      currentSCC[copyOfElement] = currentKing;
   }
   return currentKing;
} //getComponent
```

2. The method which joins the sets of two given elements (vertices)

```cpp
void uniteComponents (int firstElement, int secondElement) {
   int firstComponent = getComponent(firstElement);
   int secondComponent = getComponent(secondElement);
   if (firstComponent == secondComponent) {
      return;
   }
   if (heightOfSCC[firstComponent] < heightOfSCC[secondComponent]) {
      swap (firstComponent, secondComponent);
   }
   if (heightOfSCC[firstComponent] == heightOfSCC[secondComponent]) {
      ++heightOfSCC[firstComponent];
   }
   currentSCC[secondComponent] = firstComponent;
   minLevelInDFS[firstComponent] = min (minLevelInDFS[firstComponent], minLevelInDFS[secondComponent]);
} //uniteComponents
```

3. DFS method

```cpp
void solveIt (int node, int currentLevel) {
   int componentOfNode = getComponent(node);
   minLevelInDFS[componentOfNode] = min (minLevelInDFS[componentOfNode], currentLevel);
   visited[node] = true;
   for (auto currentNeighbour : graph[node]) {
      if (visited[currentNeighbour] == false) {
         solveIt(currentNeighbour, currentLevel + 1);
      }
      int componentOfNeighbour = getComponent(currentNeighbour);
      if (minLevelInDFS[componentOfNeighbour] < currentLevel) {
         uniteComponents(node, currentNeighbour);
      }
   }
   componentOfNode = getComponent(node);
   if (minLevelInDFS[componentOfNode] == currentLevel) {
      minLevelInDFS[componentOfNode] = MAXLevel;
   }
} //solveIt
```

4. Main method

```cpp
int main() {
   ofstream fout ("output");
   int numberOfNodes, numberOfEdges;
   readIt(numberOfNodes, numberOfEdges);
   prepareIt(numberOfNodes);
   for (int currentNode = 1; currentNode <= numberOfNodes; ++currentNode) {
      if (visited[currentNode] == false) {
         solveIt(currentNode, 1);
      }
   }
   int howmanySCC = 0;
   for (int currentNode = 1; currentNode <= numberOfNodes; ++currentNode) {
      if (currentNode == currentSCC[currentNode]) {
         ++howmanySCC;
      }
   }
   fout << howmanySCC << '\n';
   int currentIndexSCC = 0;
   for (int currentNode = 1; currentNode <= numberOfNodes; ++currentNode) {
      int currentSCC = getComponent(currentNode);
```

```cpp
      if (buildFinalSCC[currentSCC]) {
         SCC[buildFinalSCC[currentSCC]].push_back(currentNode);
      }
      else {
         ++currentIndexSCC;
         buildFinalSCC[currentSCC] = currentIndexSCC;
         SCC[buildFinalSCC[currentSCC]].push_back(currentNode);
      }
   }
   for (int currentSCC = 1; currentSCC <= howmanySCC; ++currentSCC) {
      for (auto currentNode : SCC[currentSCC]) {
         fout << currentNode << ' ';
      }
      fout << '\n';
   }
   return 0;
} //main
```

## 5. Declarations

```cpp
//Declarations - begin
const int MAXN = 1e5 + 5;
const int MAXLevel = 1e8;
vector <int> graph[MAXN];
vector <int> SCC[MAXN];
int currentSCC[MAXN];
int heightOfSCC[MAXN];
int minLevelInDFS[MAXN];
int buildFinalSCC[MAXN];
bool visited[MAXN];
//Declarations - end
```

## 6. Disjoint-set Data Structure

Disjoint-set is a data structure used for keeping disjoint sets and it was first described by Bernard A. Galler and Michael J. Fischer in 1964. It allows us to find the set which a specific element belongs to or to join the sets of two specific elements in complexity O (log* (numberOfElementsInTheSets)). The main idea is to represent each set as a tree, where the root is the father of the whole set. When two sets are to be joined, the set with the higher tree is considered the dominant one and its father becomes the father for the both sets, and when we need to find the father of a specific element, we will change the fathers of all elements on the path between it and the root, linking each of those to the root. These two optimizations generate the complexity mentioned above. More information about it may be found following the links in bibliography.

## 7. DFS Traversal

Depth-First Search is an algorithm for traversing trees or graphs (directed and undirected). A version of it was developed by French mathematician Charles Pierre Tremaux in the 19$^{th}$ century. It makes use of a stack, the top of it being considered the current node. The idea is to iterate through the neighbors of the current node and, if one of those has not been visited yet, it will be pushed in the stack and it will be set as the current node. More information about it may be found following the links in bibliography.